# High-speed silicon microring modulator at 2-μm waveband

**Weihong Shen[1,3], Gangqiang Zhou[1,3], Jiangbing Du[1,*], Linjie Zhou[1], Ke Xu[2], and Zuyuan He[1]**

[1]State Key Laboratory of Advanced Optical Communication Systems and Networks, Shanghai Jiao Tong University, Shanghai 200240, China.

[2]Department of Electronic and Information Engineering, Harbin Institute of Technology (Shenzhen), Shenzhen 518055, China.

[3]These authors contributed equally to this work.

*dujiangbing@sjtu.edu.cn

**Abstract:** We demonstrated a silicon integrated microring modulator working at the 2-μm waveband with an L-shaped PN junction. 15-GHz 3-dB electro-optic bandwidth and <1-V·cm modulation efficiency for 45-Gbps NRZ-OOK signaling is achieved at 1960 nm.

## 1. Introduction

In recent years, in virtue of hollow-core fiber [1] with low loss and low latency, and thulium-doped fiber amplifier (TDFA) [2] with broadband gain, 2-μm waveband has been considered as a promising communication window, especially for the short-reach optical interconnection and capacity/latency-hungry scenarios like data centers [3-4].

Photonic integrated circuit provides a widely-used platform for versatile devices thanks to the compact footprint and low power consumption. Because of the acceptable absorption loss of silicon at 2-μm wavelength, conventional silicon-on-insulator (SOI) is still a preferred platform for 2-μm wavelength integrated devices/systems [5]. Electro-optic modulators, which play key roles in optical communication systems, are also in urgent need to be extended towards 2 μm. Besides, weaker two-photon absorption and stronger free-carrier plasma effect make it more attractive for 2-μm-wavelength silicon modulators. Till now, several silicon modulators at the 2-μm wavelength have been reported, such as 50-Gbps Mach-Zehnder modulator (MZM) [6] and 3-Gbps microring modulator [7-8]. However, the bandwidth and data rates are still insufficient for high-speed 2-μm-wavelength transmission. Especially for the microring modulator, the current modulation bandwidth and efficiency have a large space to improve.

In this work, we report a silicon integrated microring modulator (MRM) working at the 2-μm waveband, with an L-shaped PN junction to improve the modulation efficiency [9]. The proposed MRM shows >15 GHz 3-dB electro-optic bandwidth at 2V reverse bias. The resonance shift under reverse bias is 52.5 pm/V at 1960 nm, and the corresponding modulation efficiency is 0.975V·cm. Up to 45-Gbps none-return-to-zero on-off-keying (NRZ-OOK) modulation is realized with a bit error rate (BER) under 3.8e-3. This work fills the blank space of high-speed 2-μm wavelength MRM and paves the way for high-speed silicon integrated transceivers at the 2-μm waveband.

## 2. Design and fabrication

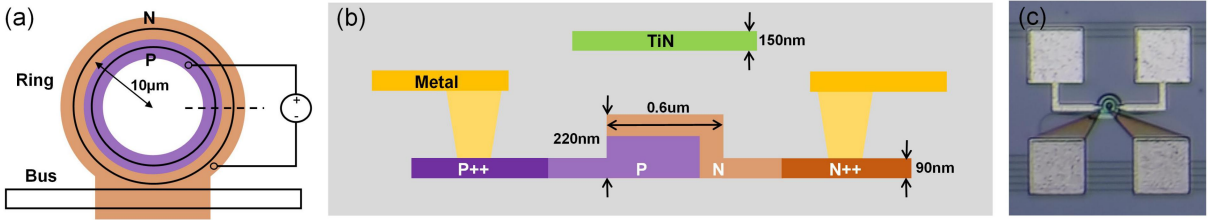

Fig. 1. (a) Microring modulator structure with the N & P implantation regions highlighted. (b) Cross-sectional schematic of the phase shifter in the MRM. (c) Micrograph of the fabricated 2-μm-wavelength MRM.

The silicon 2-μm-wavelength MRM is designed on the silicon-on-insulator (SOI) wafer with a 220-nm thick silicon layer and a 2-μm-thick buried oxide layer. The rib waveguide with a 90-nm-thick slab is 600-nm wide to support light propagation at 2 μm wavelength. As depicted in Fig. 1(b), the phase shifter in MRM has an L-shaped PN junction. The overlap between the depletion region and optical mode is larger than conventional lateral PN junction to increase the modulation efficiency. As shown in Fig. 1(a), the whole ring waveguide is integrated with a PN junction to increase the wavelength shift efficiency of the modulator when a fixed bias voltage is applied to the PN junction. The doping concentrations are 4e18 $cm^{-3}$, 1.5e18 $cm^{-3}$, 1e20 $cm^{-3}$, and 1e20 $cm^{-3}$ for the n, p, $n^{++}$, and $p^{++}$ doping regions, respectively. The highly doped regions are 0.5-μm-away from the waveguide. The PN junction is connected out by aluminum metal lines and the gap between the metal strips is 10 μm. We use a 150-nm-thick TiN thermal phase shifter to change the resonance wavelength of the MRM. The device was fabricated in AMF, Singapore. The microscope image of the fabricated 2-μm-wavelength MRM is shown in Fig. 1(c). Fiber-to-chip coupling via inverse tapers has ~6 dB/facet loss at the 2 μm wavelength. A fiber laser source with a fixed single wavelength near 1960 nm and the ASE broadband source centered at 1850 nm were employed to measure the transmission spectrum of MRM. Due to the lack of a tunable laser at 2 μm, the operation point of MRM was tuned by applying a voltage to the heater and shifting the resonance close to the laser wavelength.

## 3. Statistic measurement

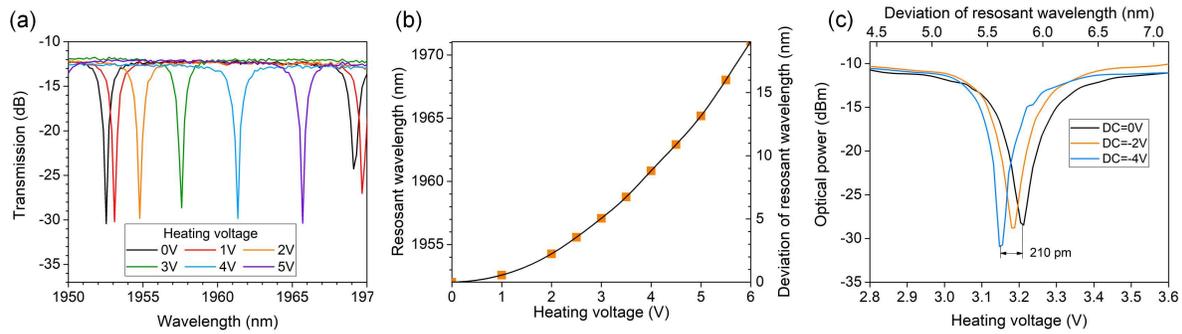

Fig. 2. (a) Measured optical spectrum of the 2-μm-wavelength MRM under 0-5V voltages on the heater. (b) Resonant wavelength and shift under 0-6V voltage on the heater. (c) Optical power after the MZM changes as a function of voltage on the heater at a fixed laser wavelength (near 1959 nm). Three bias voltages are measured.

The resonant spectra around 1960 nm under different heating voltages are plotted in Fig. 2(a), and the corresponding resonant wavelengths and shifts are extracted in Fig. 2(b). The extinction ratio (ER) near 1960 nm is >15dB. Applying heating voltage makes the resonances red-shift. The heating efficiency gradually rises with the increasing voltage and reaches 6 nm/V under 6-V heating voltage. Hence, by fitting the curve, we obtain a map with the one-to-one relationship between the resonant wavelength and heating voltage. Due to the insufficient resolution of the optical spectrum analyzer covering the 2-μm waveband, the resonant shift under the reverse bias is hard to measure accurately, which is generally two-order-of-magnitude lower than the shift under heating. Thus, we use a single-frequency laser source near 1960 nm to measure the transmitted power when the voltage on the heater is swept. In this way, we get the resonance profile as shown in Fig. 2(c). Resonant spectra under different reverse biases are also measured, and the resonant shift is 52.5 pm/V. The MRM has a 10-μm radius and a free spectrum range (FSR) of 16.29 nm. The modulation efficiency is 0.975 V·cm, which is improved by more than 4 times than the previous results in [8].

## 4. High-speed verification

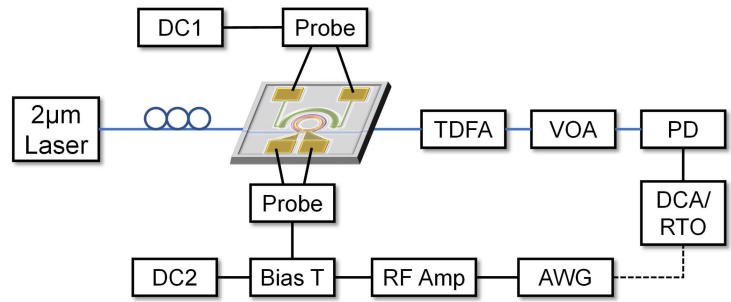

Fig. 3 High-speed RF measurement set-up for 2-μm MRM.

To verify the electro-optic (EO) response at high frequency, we set up the high-speed RF measurement for the 2-μm-wavelength MRM. The laser source is near 1960 nm with 10-dBm output power. High-speed signals are generated by an arbitrary waveform generator (AWG) with a sampling rate of 64 GSa/s and amplified to ~2.5 $V_{pp}$. The RF signal is combined with -2V DC bias by a bias-tee and loaded onto the MRM. A DC voltage is applied to adjust the proper operation point of MRM. The insertion loss of MRM is ~8dB. The modulated optical signal is amplified via TDFA and then received by a high-speed photodetector and a real-time oscilloscope for off-line digital signal processing (DSP) or a digital communication analyzer for eye-diagram measurement.

The frequency responses (S21) under various reverse biases are measured by the Agilent 67GHz vector network analyzer (VNA), as plotted in Fig. 4(a). The 3-dB EO bandwidth is beyond 15 GHz at -2V

DC bias, which is the record of high-speed MRMs at 2 μm wavelength. In the high-speed verification, a pseudorandom binary sequence (PRBS) signal was generated by the AWG with pre-equalization in order to compromise the limited system bandwidth. Eye diagrams of 20~35 Gbps on-off keying (OOK) signals were observed on the sampling oscilloscope, as shown in Fig. 4(b). 20-Gbps and 25-Gbps OOK signals present clear open eyes with a signal-to-noise ratio (SNR) of 4.5 dB and 4.3 dB. 30-Gbps OOK signal also presents a clear eye, while the eye of the 35-GHz OOK signal is almost close to the noise. A higher data rate can be achieved by applying a root raised cosine filter and feed-forward equalization (FFE) at the transmitter/receiver side. The BER curves of the 40-Gbps and 45-Gbps OOK signals are plotted in Fig. 4(c). The 40-Gbps OOK signal has a BER under 7% forward error correction (FEC) threshold (3.8e-3) at >0 dBm received optical power, and the 45-Gbps OOK signal shows a BER of 8.85e-4 at 3 dBm. The corresponding post-FFE eye diagrams are also depicted.

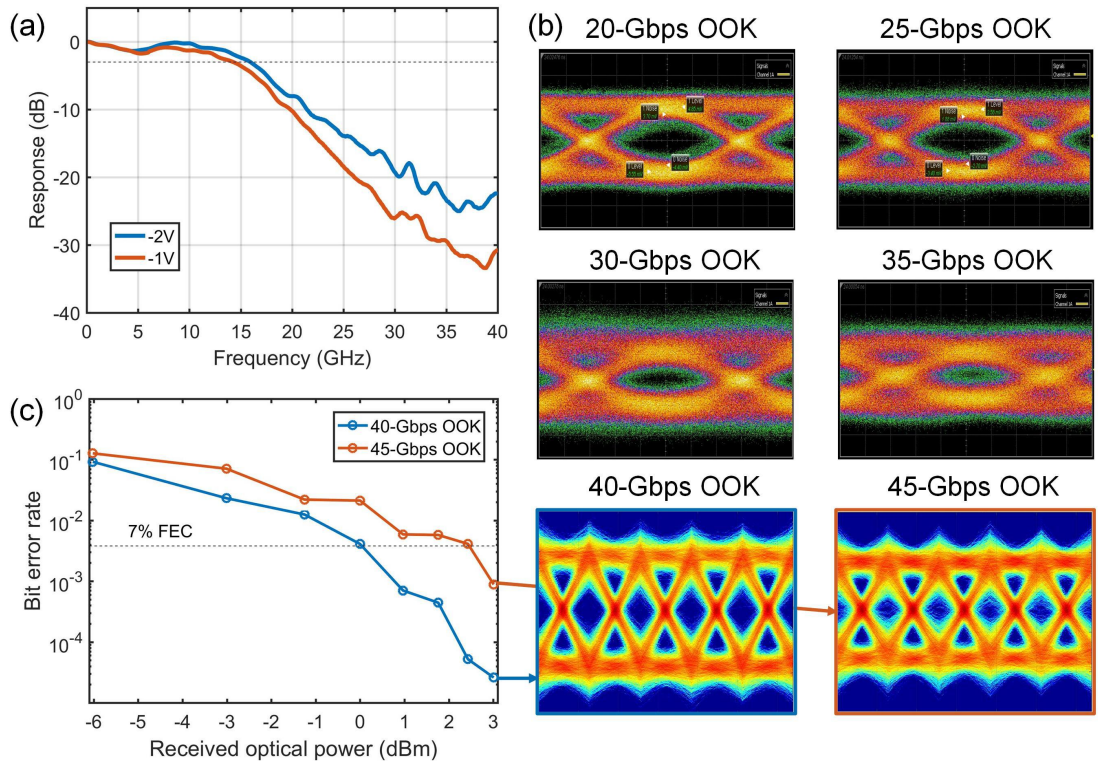

Fig. 4. (a) Electro-optic S21 responses of 2-μm-wavelength MRM under different reverse biases. (b) Eye diagrams of 20~35 Gbps OOK signals on the sampling oscilloscope. (c) BER curves of 40-Gbps and 45-Gbps OOK signals and the off-line post-FFE eye diagrams with the lowest BERs.

## 5. Conclusion

In summary, we experimentally demonstrated a silicon microring modulator working at the 2-μm waveband with a record EO bandwidth of 15 GHz and modulation efficiency of 0.975 V·cm. 45-Gbps NRZ-OOK modulation was successfully realized with a BER beneath 7% FEC threshold. This work paves the way for high-speed and high-efficiency silicon modulators at the 2-μm waveband.